\documentclass[pdfTeX]{ptephy_v1}

\preprintnumber{XXXX-XXXX} 

\usepackage{dcolumn}
\usepackage{bm}
\usepackage{color}
\usepackage{comment}
\usepackage{here}

\usepackage{lineno}

\begin{document}

\title{Polarized proton spin filter for epithermal neutron based on dynamic nuclear polarization using photo-excited triplet electron spins}


\author[1,2]{Shusuke~Takada$^{*,}$}
\author[1,3]{Kenichiro~Tateishi}%
\author[4]{Yasuo~Wakabayashi}
\author[4]{Yoshimasa~Ikeda}
\author[5]{Tamaki~Yoshioka}
\author[4]{Yoshie~Otake}
\author[1,3]{Tomohiro~Uesaka}

\affil[1]{Cluster for Pioneering Research, RIKEN, 2-1 Hirosawa, Wako, Saitama 351-0198, Japan}
\affil[2]{Department of Physics, Kyushu University, 744 Motooka, Nishi, Fukuoka 819-0395, Japan\email{takada@epp.phys.kyushu-u.ac.jp}}
\affil[3]{Nishina Center for Accelerator-Based Science, RIKEN, 2-1 Hirosawa, Wako, Saitama 351-0198, Japan}

\affil[4]{Center for Advanced Photonics, RIKEN, Wako, Saitama 351-0198, Japan}

\affil[5]{Research Center for Advanced Particle Physics, Kyushu University, 744 Motooka, Nishi, Fukuoka 819-0395, Japan}

\begin{abstract}%
For the polarization of neutrons with an \textcolor{black}{energy} of \textcolor{black}{beyond 0.1~eV}, we developed a novel polarized proton spin filter based on dynamic nuclear polarization using photo-excited triplet electron spins.
The spin filter consists of a single crystal of naphthalene doped with deuterated pentacene and has a size of $\phi15\times4$~${\rm mm}^3$, allowing it to cover a wide beam diameter. It was operated in 0.35~T and at 90~K. 
We succeeded in polarizing neutrons in the energy range $0.1-10$~eV using a RIKEN accelerator-driven compact neutron source. 
The averaged values of the proton and neutron polarization were $0.250\pm0.050$ and $0.076\pm0.015$, respectively.
\end{abstract}

\subjectindex{G12, C03}

\maketitle

\section{Introduction} \label{sec:introduction} 
Polarized neutrons have a wide range of values in physics and in industry. 
Notably, the polarized epithermal neutrons whose \textcolor{black}{energy} is in the range $0.1-1,000$~eV are useful mainly for studies on neutron-induced compound states which are formed by the neutron capture when the epithermal neutron beam is injected into a nucleus target.
One of the studies using the compound states involves a search for time-reversal (T) violation, which is an open question in elemental particle physics for explaining the development of the universe.
T-violation search with the compound states can allow studies for new physics beyond the standard model of elementary particles, e.g., supersymmetry~\cite{gudkov2017pseudomagnetic, gudkov1992cp, okudaira2018angular}. 
In the compound states, the extremely large helicity dependence of the capture cross-sections has been observed in various nuclei \cite{mitchell2001parity}. 
The helicity dependences have proven the existence of the parity (P) violations in the compound states, which are $10^6$ times larger than that of the proton-proton scattering.
The large P-violation is theoretically explained as an interference between s- and p-wave resonances. Moreover, the T-violation with the compound states may also be enhanced with a similar mechanism \cite{gudkov1992cp}.
T-violation search will be completed with the polarized epithermal neutron beam and a polarized nucleus target. 

Neutron spin filters are devices for neutron beam polarization, which rely on the spin-dependent cross-section of nuclear capture on polarized $^3$He nuclei or scattering on polarized protons.
The former is now available at several neutron facilities; however, it is impossible to optimize a filter size for the wide-energy range because the capture cross-section decreases rapidly in the epithermal region.
The latter is currently the only idea for neutron polarization with energies up to keV, because neutron-proton scattering cross-section is nearly constant in the epithermal region.
The polarized proton spin filter was firstly demonstrated by Lushchikov $et~al.$ using a method of dynamic nuclear polarization (DNP) \cite{luschikov1969polarized}.
In this method, electron polarization is transferred to a proton via microwave irradiation for polarizing proton spins in solids \cite{jeffries1963dynamic, abragam1982nuclear}.
DNP relies on the thermally equilibrated high polarization of electrons, which is added to samples as polarizing agents and is realized at cryogenic temperatures ($\sim$1~K) and with a strong magnetic field (2.5-5~T).
In addition to the foregoing strict environment, because the flux of the thermal neutron (meV-eV) is higher than that of the epithermal neutron in most pulsed neutron sources, the polarized $^{3}$He spin filter has become preferred over the polarized proton spin filter.

Recently, the polarized proton spin filter has been demanded again because fluxes of the epithermal neutrons are increasing at neutron facilities, e.g., Japan Proton Accelerator Research Complex (J-PARC).
In the meantime, an alternative method for polarizing protons has been developed.
This method is called DNP with photo-excited triplet electron spins (triplet DNP), wherein non-equilibrated electron spins are utilized.
The selection rule determines the polarization of the triplet electron in the intersystem crossing, which is independent of temperature and magnetic field strength.
The neutron polarization is realized with the simple setup because of the milder environment compared to that of conventional DNP, so that the polarized proton spin filter based on triplet DNP (triplet-DNP spin filter) can be installed at existing beamlines.
Since a leakage magnetic field is small due to the milder field, the experimental equipment, e.g., nucleus target and detector, can be set near to each other in a beamline. 
\textcolor{black}{Since the spin filter and the detector can be placed close together, we can use the neutron beam before diverging it.}
The pioneering work of the triplet-DNP spin filter was firstly conducted at the Paul Scherrer Institut (PSI) in Switzerland \cite{haag2012spin, eichhorn2013apparatus, eichhorn2014proton, quan2019transportable}.
They achieved a proton polarization of 0.80 at 25~K and in 0.36~T, using a single crystal of naphthalene doped with deuterated pentacene with a size of $5\times5\times5$~${\rm mm}^3$.
They applied it to the cold neutron and carried out a small-angle neutron scattering (SANS) experiment \cite{niketic2015polarization}. 
\textcolor{black}{However, the triplet-DNP spin filter has never been applied to epithermal neutrons.}

This paper reports the first demonstration of the polarization of the epithermal neutrons with the triplet-DNP spin filter.
Firstly, the working principle of the polarized proton spin filters is reviewed in Section~\ref{sec:principle_spinfilter}.
The optimal thickness of the triplet-DNP spin filter for the epithermal region is likewise shown therein. 
In Section~\ref{sec:rans_experiment}, \textcolor{black}{we describe an experimental setup of a neutron transmission using the triplet-DNP spin filter and a neutron beam.}
Considering the triplet-DNP spin filter, the single crystal of naphthalene doped with deuterated pentacene was used with the size of $\phi15\times4$~${\rm mm}^3$. 
Thereafter, \textcolor{black}{the} triplet DNP was carried out at 90~K and in 0.35~T.
The performance of the triplet-DNP spin filter was evaluated by comparing the neutrons that pass through the triplet-DNP spin filter with and without proton polarization.
The evaluation of the performance is described in Section~\ref{sec:result_discussion}.
Section \ref{sec:result_discussion} likewise discusses the comparison with the polarized $^{3}$He spin filter and future improvements thereon.
The neutron transmission experiment was conducted using the RIKEN accelerator-driven compact neutron source (RANS) \cite{otake2017research}.
RANS provides a pulsed neutron beam with a wide-energy region from meV to MeV.
Enough neutron \textcolor{black}{beam} intensity for the evaluation could be obtained by optimizing the experimental setup.
Proton polarization was kept for 70~h without radiation damage.
Neutron polarization in the epithermal region was clearly observed.


\section{Design of triplet-DNP spin filter for epithermal neutron} \label{sec:principle_spinfilter}

The triplet-DNP spin filter is based on the polarization of hydrogen nuclei which are contained in a naphthalene crystal.
The principle of neutron polarization with the polarized proton spin filter relies on the fact that the singlet cross-section for neutron-proton scattering is twenty times larger than the triplet cross-section \cite{gurevich1968low, luschikov1969polarized}.
The total cross-section of neutron-proton scattering is customarily defined as the sum of spin-dependent and -independent cross-sections ($\sigma_{\rm p}$ and $\sigma_{0}$) ~\cite{abragam1982nuclear, glattli1987methods, borner2003neutron}: 
\begin{equation}
  \sigma=\sigma_{0} + \sigma_{\rm p} P ({\bm S} \cdot {\bm I}), 
  \label{eq:sigma}
\end{equation}
where $P$ is the proton polarization. ${\bm S}$ and ${\bm I}$ are the unit vectors of an incident neutron spin and a proton spin, respectively.
Thus, the neutrons that are polarized anti-parallel to the protons will much strongly interact compared to those that are polarized parallel thereto.
Considering that the main component of the triplet-DNP spin filter is naphthalene, an unpolarized neutron beam is exponentially attenuated by passing through it with the proton and the carbon densities $n$ and $n_{\rm C}$, respectively, and a filter thickness $d$. 
Neutron transmissions of the polarized and unpolarized proton spin filter ($T_{\rm n}$ and $T_{\rm n0}$) are expressed using the following equations: 
\begin{align}
  \label{eq:trans_unpol}
  T_{\rm n0} &= \exp{ \left\{ -\left(n \sigma_{\rm 0} + n_{\rm C} \sigma_{\rm C} \right) d\right\} }, \\
  \label{eq:trans_pol}
  T_{\rm n} &= \exp{ \left\{-\left(n \sigma_{\rm 0} + n_{\rm C} \sigma_{\rm C} \right) d \right\}} \cosh{ \left( n \sigma_{\rm p} P d \right) }, \\
  \label{eq:ratio_trans}
  \frac{T_{\rm n}}{T_{\rm n0}} &= \cosh{ \left(n \sigma_{\rm p} P d \right) },
\end{align}
where $\sigma_0$ and $\sigma_{\rm C}$ are the cross-sections of neutron scattering with the unpolarized proton and the carbon nucleus, respectively.
The neutron polarization $P_{\rm n}$ after passing through the spin filter is written as~\cite{gurevich1968low}
\begin{equation}
  P_{\rm n}=\tanh{ \left(n \sigma_{\rm p} P d \right)}.
  \label{eq:poln}
\end{equation}
A figure of merit (FOM) is taken as the statistically relevant factor in the optimization of the spin filter performance \cite{williams1988polarized}, and it is defined by the following equation:
\begin{equation}
  {\rm FOM}=P_{\rm n}^{2} T_{\rm n}.
  \label{eq:fom}
\end{equation}

\begin{figure}[htbp]
  \centering
  \begin{minipage}{0.75\textwidth}
    \includegraphics[keepaspectratio,scale=0.6, angle=0]{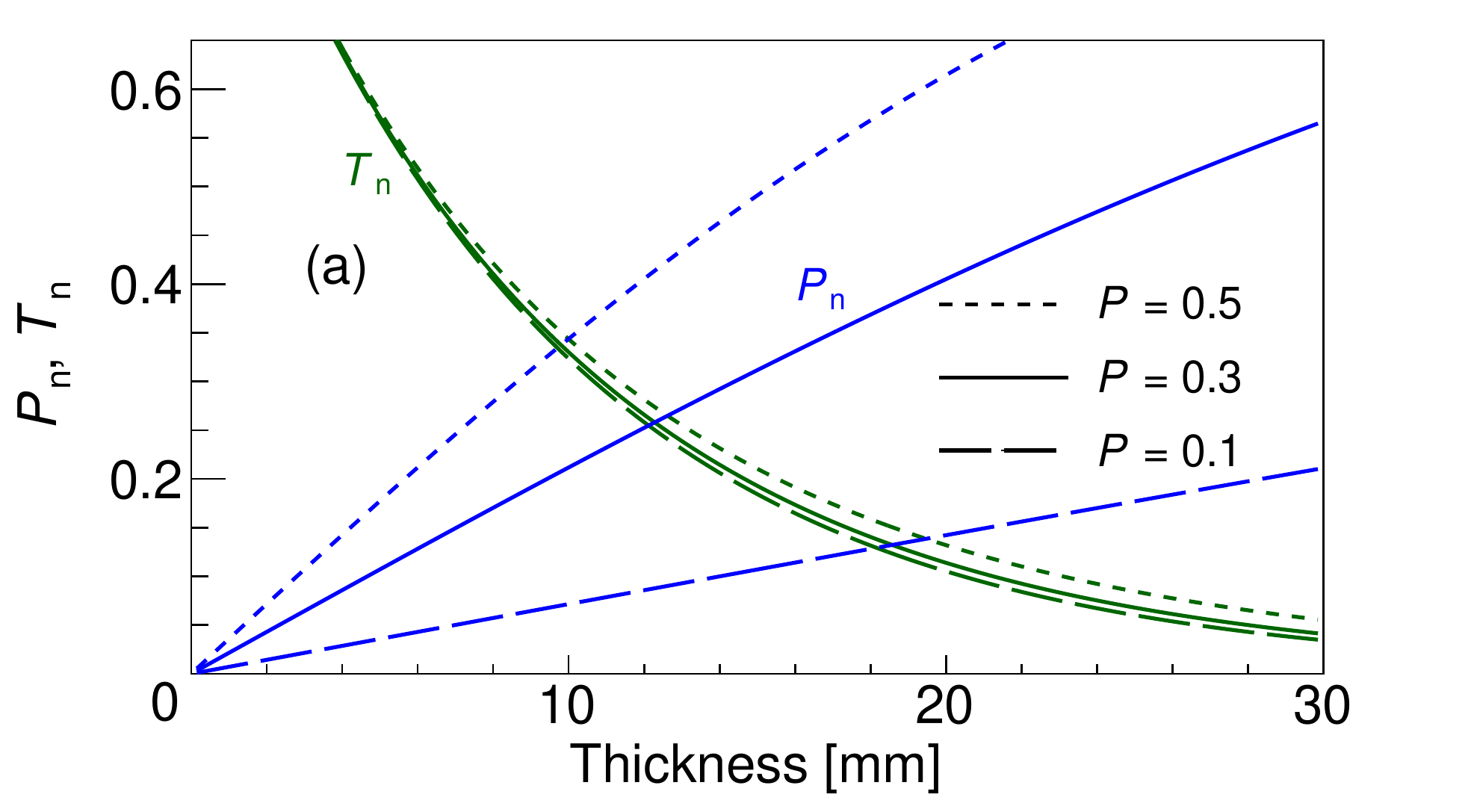}
  \end{minipage}\\
  \begin{minipage}{0.75\textwidth}
    \includegraphics[keepaspectratio, scale=0.6, angle=0]{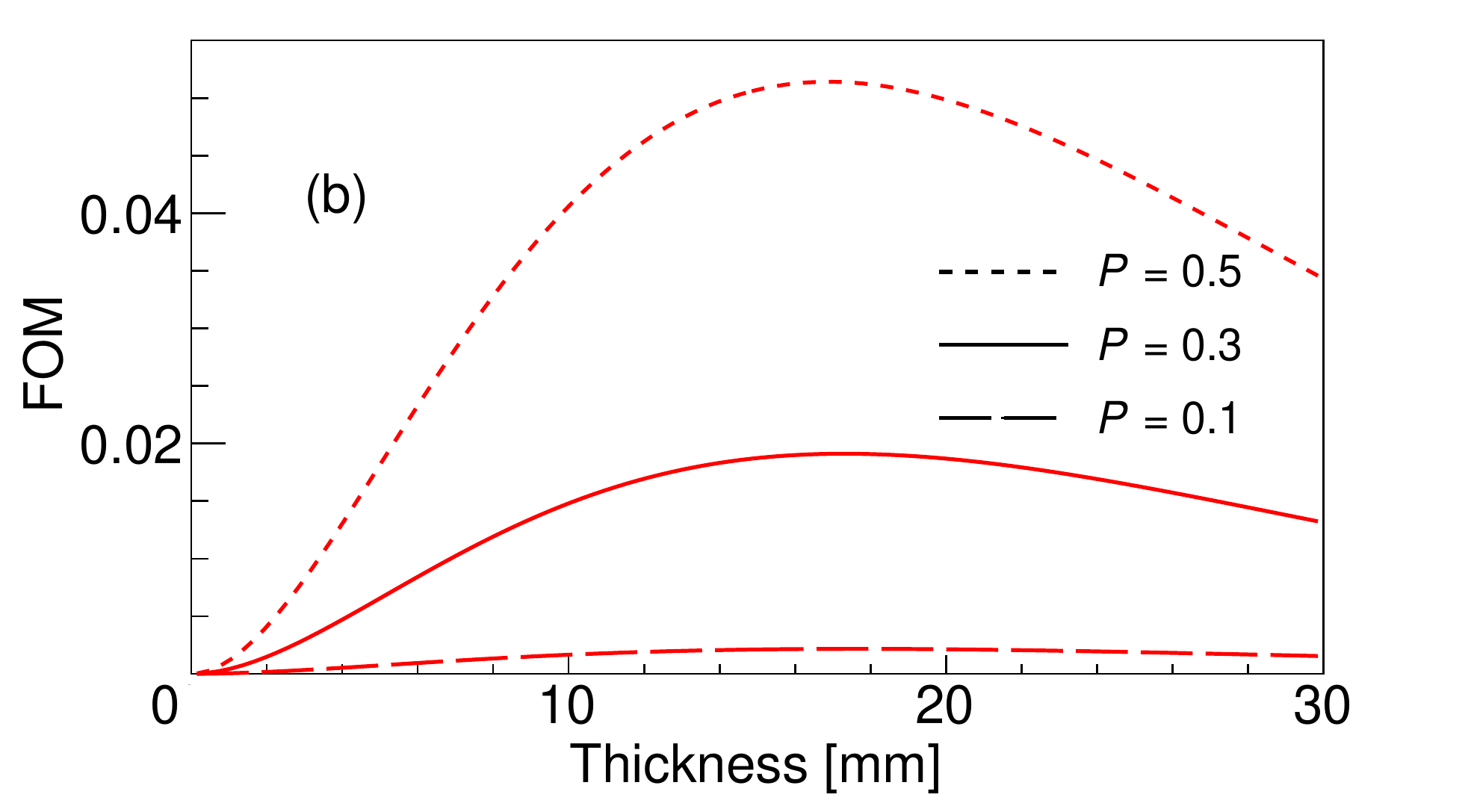}
  \end{minipage}
  \caption{Performance of the polarized proton spin filter with the proton polarization $P=0.1, 0.3$ and $0.5$ for epithermal neutron. (a) Degree of neutron polarization ($P_{\rm n}$) and neutron transmission ($T_{\rm n}$) as a function of spin filter thickness. (b) Figure of merit (FOM) as a function of spin filter thickness.}
  \label{fig:fom_thickness}
\end{figure}

The performance of the triplet-DNP spin filter in the epithermal region is calculated using Eq.~\ref{eq:fom} as a function of the filter thickness $d$. 
Fig.~\ref{fig:fom_thickness} shows the neutron polarization, the neutron transmission, and the FOM for the $P=0.1/ 0.3 /0.5$ cases. 
Here, $n_{\rm p}=4.29 \times 10^{22}$~$\rm /cm^3$, $n_{\rm C}=5.36\times10^{22}$~$\rm /cm^3$, $\sigma_{\rm 0}=20.5$~barn, $\sigma_{\rm p}=16.7$~barn, and $\sigma_{\rm C}=4.74$~barn were used.
A thicker filter leads to a higher neutron polarization and a lower transmission.
The optimum thickness is $\sim$15~mm. Moreover, it is almost independent of the proton polarization.

Triplet DNP using a single crystal of naphthalene was originally demonstrated by Henstra $et~al.$ in 1990 \cite{henstra1990high}.
This method has a substantial advantage over DNP with radicals.
It enables the production of nuclear hyperpolarization at a relatively lower magnetic field and a higher temperature.
This can reduce a stray magnetic field as well as make expensive and high-tech cryogenic devices unnecessary.
The proton polarization of 0.80 was obtained at 25~K and in 0.36~T \cite{quan2019transportable}.
That of 0.34 was achieved even at room temperature and in 0.40~T \cite{tateishi2014room}.
Recently, this method was applied not only in accelerator sciences but also in the chemical and medical fields \cite{tateishi2013dynamic, negoro2018dissolution, fujiwara2018dynamic}.

\begin{figure}[htbp]
  \centering
    \includegraphics[scale=0.7]{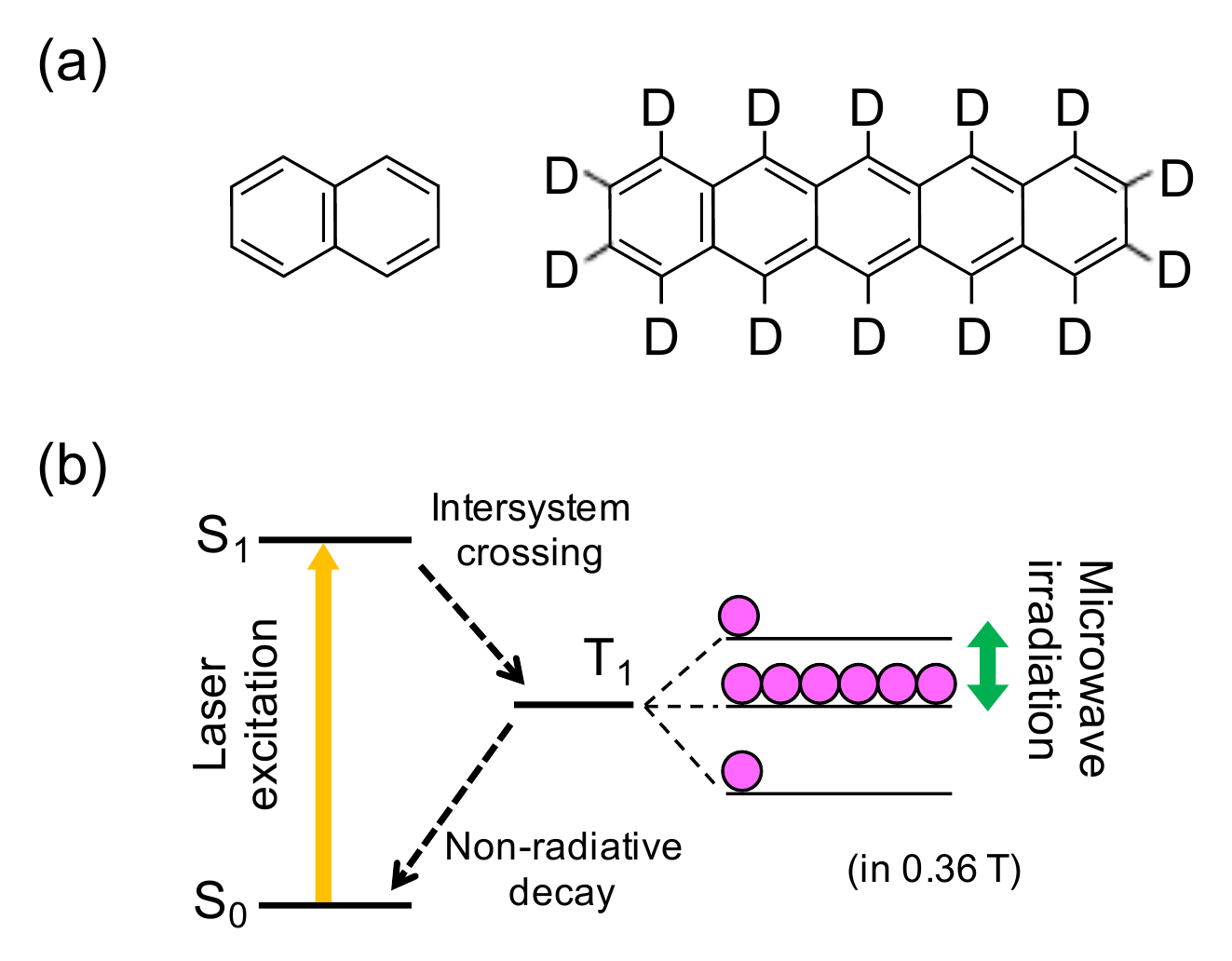}
  \caption{(a) Naphthalene (left) and deuterated pentacene (right) molecule. (b) Simplified energy diagram of pentacene showing a pathway during triplet DNP.}
  \label{fig:pentacene}
\end{figure}

\begin{figure}[htbp]
  \centering
    \includegraphics[scale=0.8]{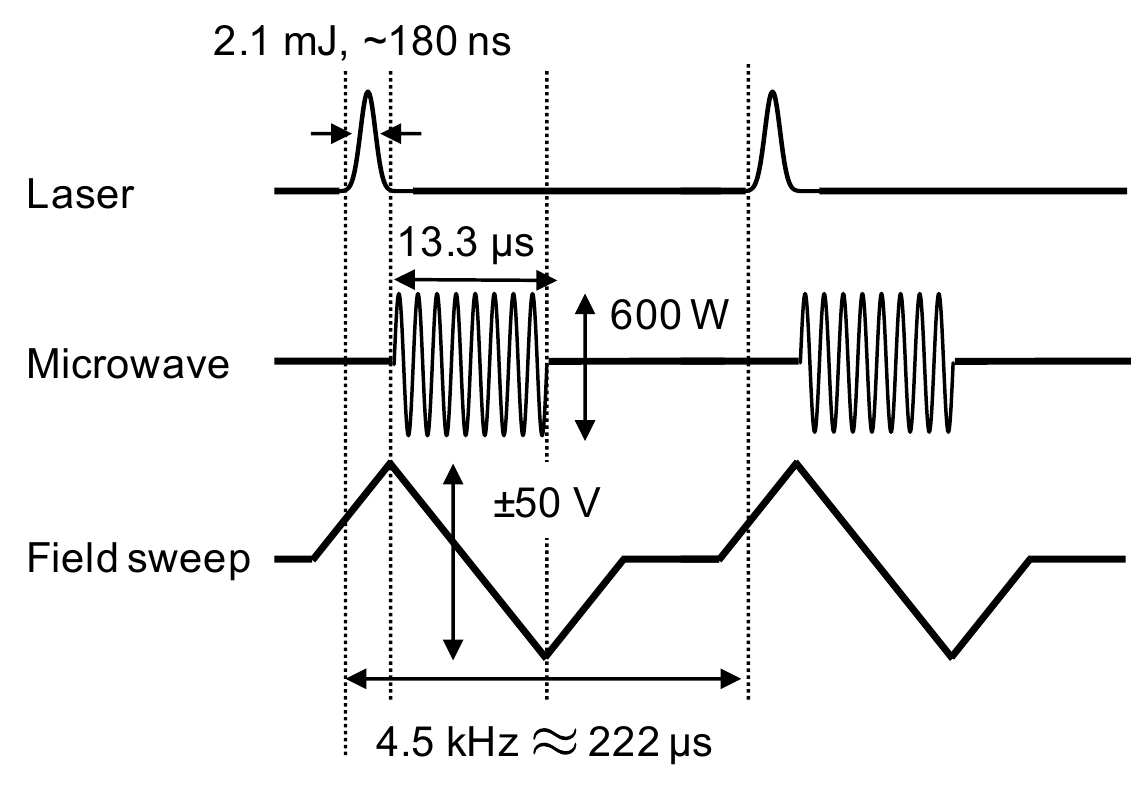}
  \caption{Pulse sequence of triplet DNP. One cycle is composed of a laser pulse for generating a hyperpolarized triplet electron and a microwave pulse and a field sweep for polarization transfer.}
  \label{fig:pulse_sequence}
\end{figure}

We applied triplet DNP to a single crystal of naphthalene doped with 0.003~mol\% deuterated pentacene (Fig.~\ref{fig:pentacene}(a)). 
The polarization procedure of \textcolor{black}{the} triplet DNP, which is explained in Ref. \cite{eichhorn2014dynamic}, begins with laser irradiation to generate hyperpolarized electrons
Fig.~\ref{fig:pentacene}(b) shows the energy diagram of pentacene.
Light irradiation with the wavelength of 589~nm induces the transition from its ground singlet state ${\rm S}_0$ to its excited triplet state ${\rm T}_1$ via its excited singlet state ${\rm S}_1$.
The transition probability determines the population of the triplet state. 
In the case of pentacene, the populations are 0.045, 0.910, and 0.045, which corresponds to the electron polarization of 0.906 when the long axis of the molecule is aligned parallel to the external magnetic field \cite{van1980epr}.
The hyperpolarization of triplet electrons is transferred to nearby protons during the lifetime of the electron through the process called the integrated solid effect (ISE) \cite{henstra2014dynamic}.
Fig.~\ref{fig:pulse_sequence} shows a cycle of the ISE.
In the ISE, a magnetic field sweep and microwave irradiation are applied simultaneously.
The inhomogeneously broadened electron spin packets are swept adiabatically. 
We found that the Rabi frequency of the electron spin in the rotating system matches the Larmor frequency of proton spin at some point in the adiabatic process.
The excited electrons decay non-radiatively to ${\rm S}_0$ and the hyperpolarized spin state diffuses to the whole naphthalene crystal.
By repeating this cycle, the proton polarization can be accumulated until the buildup and proton spin-lattice relaxation are balanced.

Triplet DNP was conducted in 0.35~T using a C-type electromagnet with a gap of 100~mm and a pole diameter of 220~mm.
The resonance frequencies of electron and proton were 9.2~GHz and 15.0~MHz (high-field transition), respectively.
A diode-pumped solid-state (DPSS) laser (CNI, HPL-589-Q) was used for pentacene excitation.
The wavelength, pulse width, pulse energy, and repetition rate were 589~nm, $\sim$180~ns, 2.1~mJ, and 4.5~kHz, respectively.
The laser pulses were sampled with a photodiode and converted to TTL level. 
The signals were used as a trigger for the subsequent microwave and field sweep.
A microwave pulse amplified up to 600 W using pulsed TWTA (IFI, PT188-1KW, max duty 6\%), with a width of 13.3~${\rm \mu}$s, was applied, while the field was adiabatically swept with the voltage of $\pm$50~V.
The transmission loss was around -3~dB.
The polarized proton signals were monitored using the OPENCORE NMR spectrometer \cite{takeda2007highly, takeda2008opencore}.

\begin{figure}[htbp]
  \centering
    \includegraphics[scale=.8]{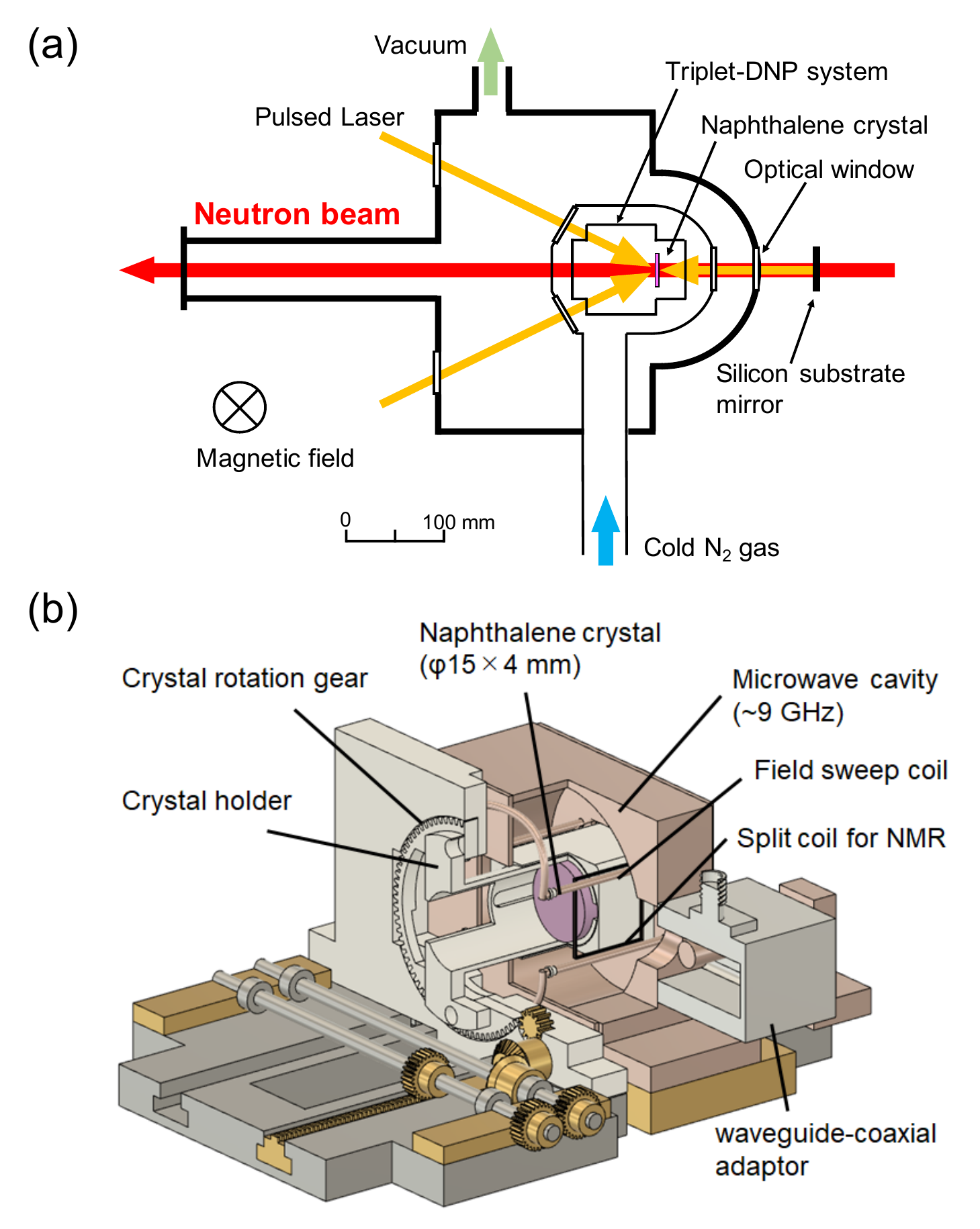}
  \caption{(a) Experimental setup of triplet DNP. A double-walled chamber is put in an electromagnet. The inner chamber was cooled down to 90~K. and the outer chamber was kept at $\sim$20~Pa. Laser lights were irradiated from three lines. (b) A schematic drawing of the home-built ${\rm TE}_{011}$ cylindrical cavity equipped with a field-sweep coil and a split-coil for NMR. The naphthalene crystal can be rotated using a crystal rotation gear.}
  \label{fig:triplet_system}
\end{figure}

A system of triplet DNP is set in a double-walled chamber (Fig.~\ref{fig:triplet_system}(a)).
The inner chamber was cooled down to 90~K using cooled nitrogen gas and the outer chamber was kept at $\sim$20~Pa to prevent frosting.
Optical windows are attached to the chambers. 
\textcolor{black}{The laser light was introduced from three places: one from the upstream and two from the downstream.} 
\textcolor{black}{For the upstream, a dielectric-coated silicon substrate mirror with the thickness of 1 mm was used to minimize the neutron transmission loss.}
The inside of the chamber is shown in Fig.~\ref{fig:triplet_system}(b). 
The naphthalene crystal was placed at the center of the electromagnet. 


The crystal was cut into the size of $\phi15\times4$~${\rm mm}^3$ because the power of our laser was insufficient to polarize the 15~mm thick filter.
It was mounted on a Teflon holder. 
A ${\rm TE}_{011}$ cylindrical cavity equipped with a field-sweep coil and a split-coil for NMR was utilized.
The diameter and length are 21~mm and 25~mm, respectively. 
A coaxial microwave transmission line was adapted to a waveguide and coupled with the cavity through an iris.
The crystal can be rotated using a crystal rotation gear. Moreover, its alignment can be adjusted precisely.

\section{Optical layout for transmission experiment} \label{sec:rans_experiment}

\begin{figure}[htbp]
  \centering
    \includegraphics[scale=0.6]{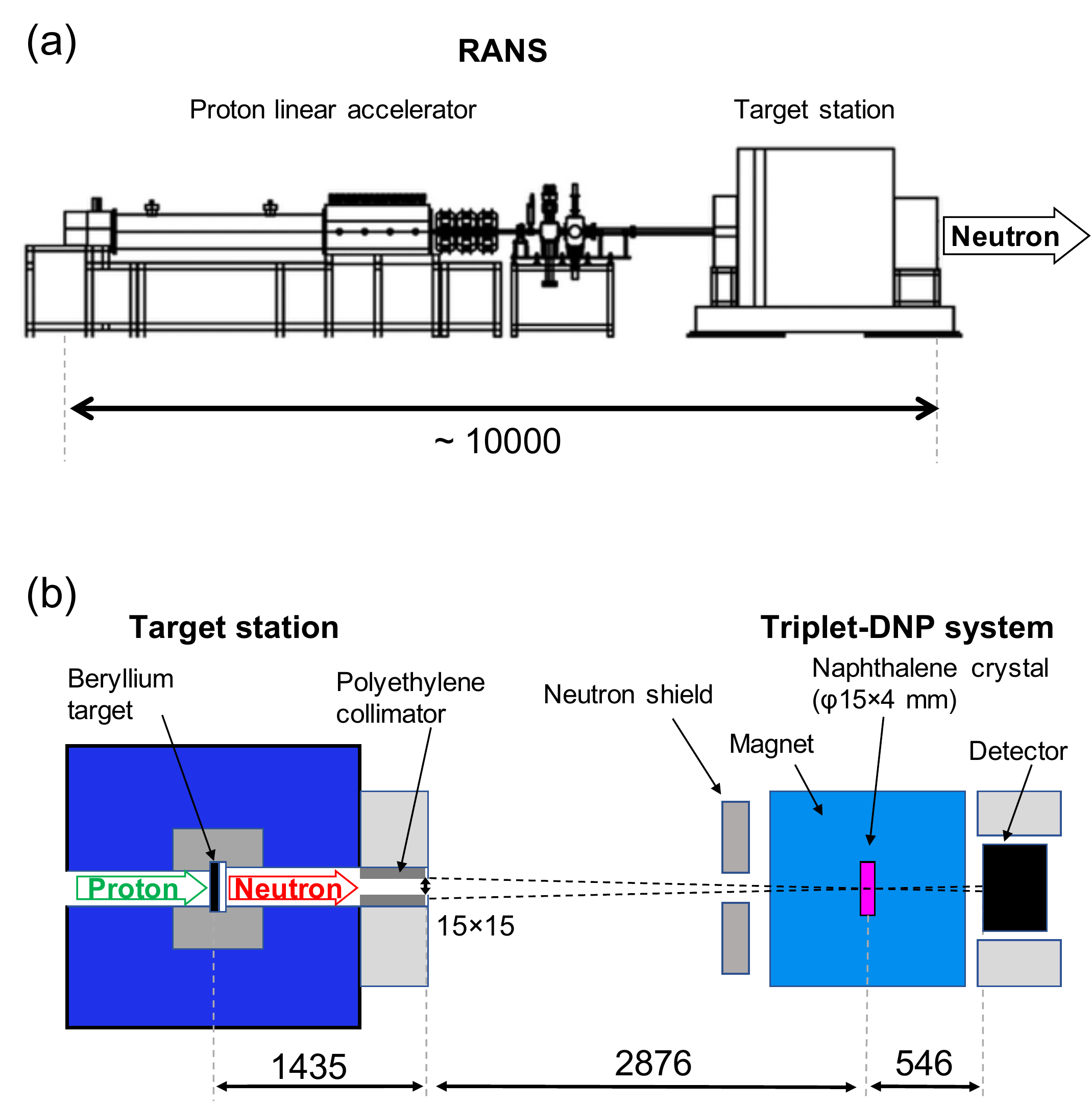}
  \caption{(a) Side view of RANS. (b) Top view of experimental setup of neutron transmission experiment with RANS.}
  \label{fig:rans_setup}
\end{figure}

The performance of the triplet-DNP spin filter was tested by measuring neutron transmission with RANS.
RANS is a compact neutron source which has been in operation since 2013.
RANS has been applied to the developments of non-destructive inspection methods for infrastructures and industrial products, e.g., concrete and steel \cite{takamura2016non, ikeda2017nondestructive, taketani2016visualization, wakabayashi2019feasibility, yoshimura2019neutron}.
RANS consists of a linear accelerator and a target station as shown in Fig.~\ref{fig:rans_setup}(a).
Protons are accelerated to 7~MeV and injected into a beryllium (Be) target \cite{takamura2016non}.
Neutrons with the maximum energy of $\sim$5~MeV are generated via the $^9$Be (p,n) reaction.
The neutrons are slowed down through a 40~mm thick polyethylene moderator and extracted from the target station.
An energy spectrum at a position which is 5~m from the Be target is shown in Ref. \cite{otake2017research}. 
The Be target and the moderator are surrounded with carbon blocks as a neutron reflector, borated polyethylene (BPE) powder in aluminum boxes, and lead blocks in iron enclosures shield the neutrons and gamma-rays.

\begin{figure}[htbp]
  \centering
    \includegraphics[scale=0.6]{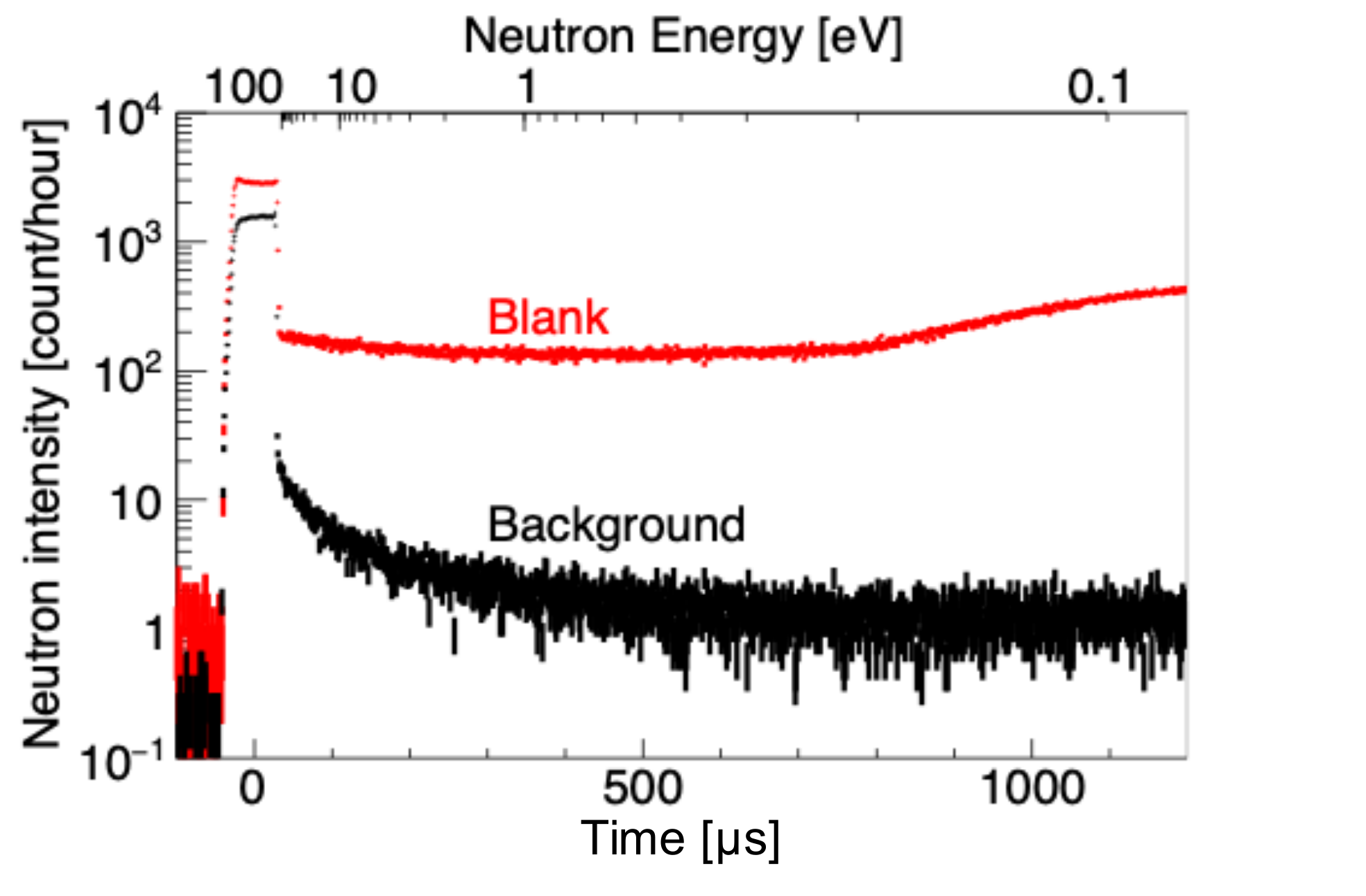}
  \caption{\textcolor{black}{Neutron \textcolor{black}{beam intensity} of blank and background measurements} in the epithermal region as a function of the time of flight.}
  \label{fig:rans_tof}
\end{figure}


The experimental setup is shown in Fig.~\ref{fig:rans_setup}(b). 
The 10\% borated polyethylene (BPE) collimator with a hole of $15\times15$~${\rm mm}^2$ was installed at the RANS target station.
A neutron shield for reducing background signal consists of BPE blocks and boron rubber sheets.
An RPMT detector, which consists of a ZnS(Li) scintillator and a position-sensitive photomultiplier tube \cite{satoh2018development}, was used for measuring a 2D position and a neutron time of flight (TOF).
The detector was set in an aluminum box with BPE powder to shield background neutrons and placed 0.55~m behind the naphthalene crystal. 
At that position, a stray field from the electromagnet was less than 10~G, and it did not affect the photomultiplier tube of the detector. 
\textcolor{black}{The spatial resolution and the neutron beam intensity are as the function of the distance from the neutron source to the spin filter. 
The longer the distance, the better the spatial resolution, but the smaller the neutron beam intensity. 
The distance set to be 4.31~m, and the resolution and the intensity in a whole energy region were roughly 2mm, and  $8.3\times10^5$~count/hour, respectively.}

A neutron pulse structure determines the maximum and minimum available neutron energy.
The shorter pulse width led to a higher maximum available neutron energy, but a smaller neutron \textcolor{black}{beam intensity}.
In this experiment, pulse width and repetition were set at 60~$\mu$s and 105~Hz, respectively. 
Considering the repetition rate and the distance from Be target to the detector, the minimum available neutron energy was 1~meV.
\textcolor{black}{We performed the measurement without a target to calculate neutron transmission (blank measurement). 
In order to subtract unexpected events, such as a scattering with a microwave cavity, we carried out the measurement with a boron rubber target (33~mm thickness) instead of the naphthalene crystal (background measurement).}
Fig.~\ref{fig:rans_tof} shows the epithermal region of the TOF \textcolor{black}{spectrums} \textcolor{black}{with the blank measurement and the background measurement}.
\textcolor{black}{The neutron \textcolor{black}{beam intensity} of the background measurement is $10-10^2$ lower than the corresponding blank measurement for all time regions in the TOF spectrum.}
Considering the pulse width and the distance from the Be target to the detector, the maximum available neutron energy was 10~eV. 
The TOF signals in a full range is shown in Supplementary Fig.~S1.
The neutron \textcolor{black}{beam intensity} has $1.7\times10^5$~count/hour in the $0.1-10$~eV range.


\section{Neutron polarization experiment} \label{sec:result_discussion} 

\begin{figure}[htbp]
  \centering
    \includegraphics[scale=0.6, angle=0]{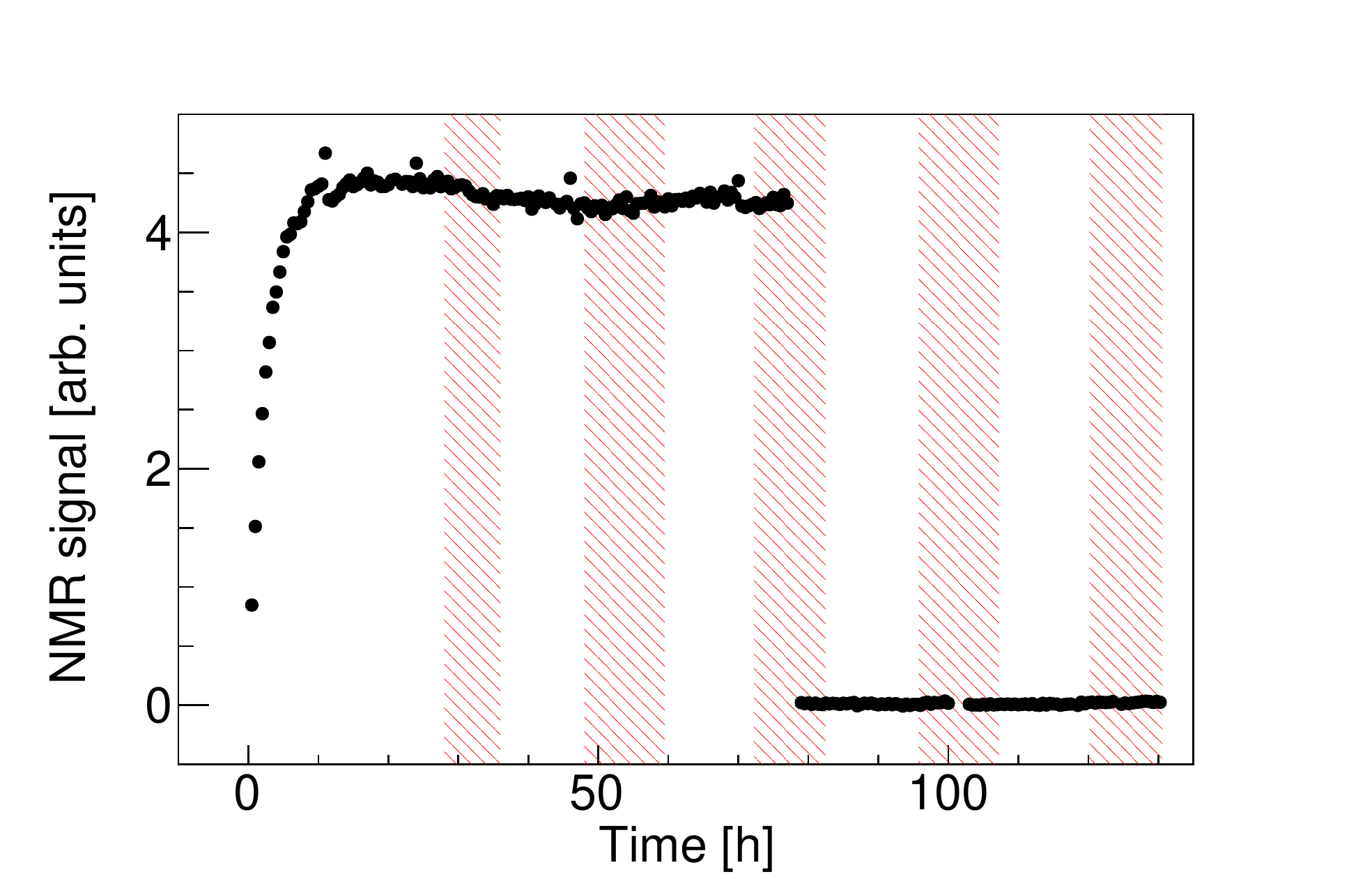}
  \caption{Relative intensity of the proton polarization during the beamtime. The shaded areas pertain to the neutron irradiations.}
  \label{fig:nmr_monitor}
\end{figure}

The relative magnitude of the proton polarization was monitored using the NMR method mentioned in Section \ref{sec:rans_experiment}. 
Fig.~\ref{fig:nmr_monitor} shows the intensities of the NMR signals recorded during the beamtime.
\textcolor{black}{The expanded figure of Fig.~\ref{fig:nmr_monitor} in the 0-15~hrs is shown in Supplementary Fig.~S2, then the buildup time was about 2.5~hrs.}
The shaded areas show the irradiation time of the neutron beam.
We irradiated the neutron beam after the buildup of the proton polarization was fully saturated. 
Radiation damage was not observed because the signal intensity did not decrease during the neutron irradiations.
The proton was depolarized at around $T = 80$~hrs, and this state was maintained. 
The depolarized state was realized by shifting the relative timing of the microwave and the field sweep to the laser \textcolor{black}{instead of cutting off} the polarization sequence shown in Fig.~\ref{fig:pulse_sequence} in order to keep the environment inside the inner chamber.
\textcolor{black}{At the offline measurement, the relaxation time in the irradiation of the laser was 5.8~hrs (shown in Supplementary Fig.~S2). 
Using the buildup constant and the relaxation time obtained from the NMR results, the proton polarization was calculated to be 51.2\%. 
However, an unpolarized area due to the lack of laser power is not included in the above estimation. 
The total proton polarization has to be measured by using neutrons.
}

\begin{figure}[htbp]
  \centering
    \includegraphics[scale=0.6, angle=0]{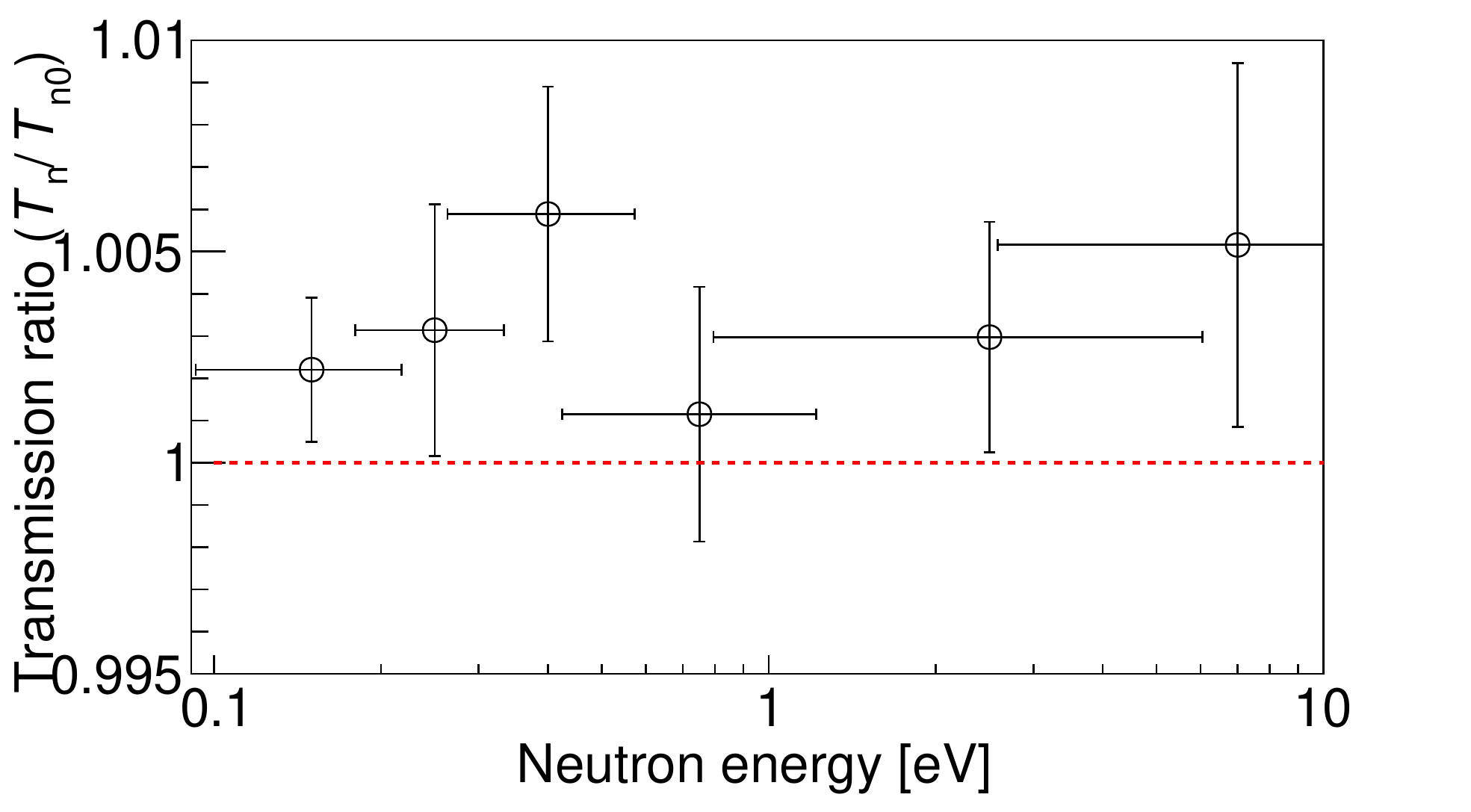}
  \caption{Ratio of neutron transmissions ($\frac{T_{\rm n}}{T_{\rm n0}}$) as a function of neutron energy.}
  \label{fig:ratio_trans}
\end{figure}

\begin{figure}[htbp]
  \centering
    \includegraphics[scale=0.7]{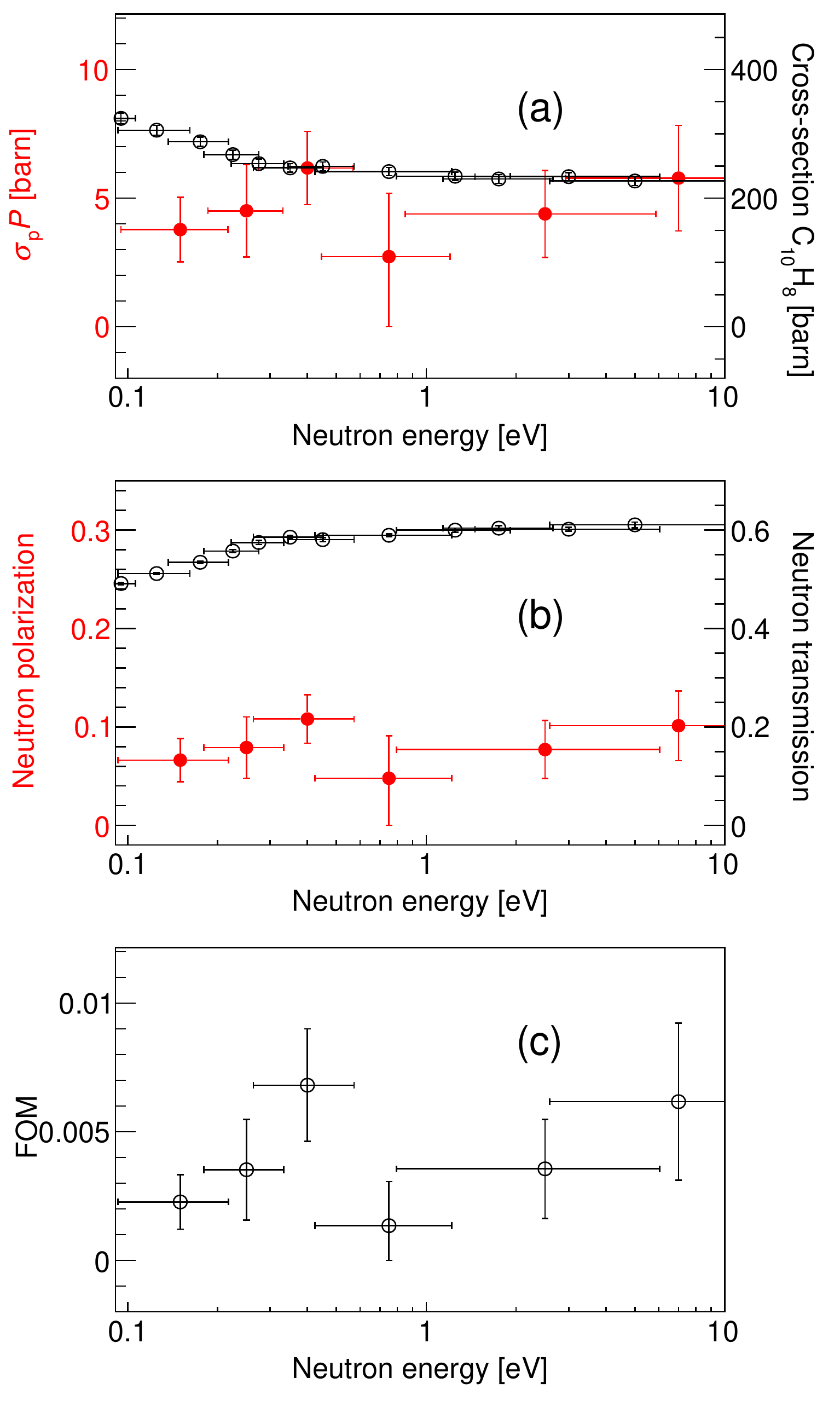}
  \caption{(a) $\sigma_{\rm p} P$ (Filled circle) and the total cross-section of naphthalene (white circle). (b) Degree of neutron polarization (filled circle) and neutron transmission (white circle). (c) Figure of merit. These results are measured at 90~K and 0.35~T.}
  \label{fig:summary}
\end{figure}

An absolute value of the proton polarization is analyzed from comparing the neutrons pass through the triplet-DNP spin filter with and without proton polarization. 
\textcolor{black}{Figure~\ref{fig:ratio_trans} shows a ratio of the TOF spectrum with and without proton polarization of the naphthalene crystal, which is consistent with the ratio of neutron transmission $T_{\rm n}/T_{\rm n0}$.}
Here, the TOF is converted to a neutron energy $E_{\rm n}$ by the following equation:
\begin{equation}
  E_{\rm n} = \frac{m_{\rm n}}{2} \left( \frac{L}{t} \right)^2
  \label{eq:TOFtoNeutronEnergy}
\end{equation}
where $m_{\rm n}$ is the neutron mass, $L$ is the distance between the Be target and the RPMT detector and $t$ is the TOF.
\textcolor{black}{The ratio of the neutron transmission $T_{\rm n}/T_{\rm n0}$ as a function of neutron energy is reasonable according to Eq.~\ref{eq:ratio_trans}} because the ratio is greater than one over the whole region, which is shown in Fig.~\ref{fig:ratio_trans}. 
\textcolor{black}{The ratio of the neutron transmission in the energy region} from 1~meV to 10~eV is shown in Supplementary Fig.~S3.
Figure~\ref{fig:summary}(a) shows $\sigma_{\rm p} P$, which is calculated from the ratio of the neutron transmission \textcolor{black}{$T_{\rm n}/T_{\rm n0}$} (Fig.~\ref{fig:ratio_trans}) and Eq.~\ref{eq:ratio_trans}. 
The error of $\sigma_{\rm p} P$ includes the statistical error of transmitted neutrons and the filter thickness of $4.1\pm0.1$~mm. 
By dividing $\sigma_{\rm p} P$ in Fig.~\ref{fig:summary}(a) by the literature value of $\sigma_{\rm p}$, the absolute value of the proton polarization is calculated.
The absolute value is only obtained in the energy region of $0.1-10$~eV because neutron-proton scattering cross-section is practically constant in \textcolor{black}{constant in the region higher than 0.1~eV}, independent of filter material and temperature \cite{granada1984total, squires2012introduction, mattes1984jef, grieger1998total, morishima2002cross}.
Thus, we applied the average value of $\sigma_{\rm p}$ at $0.1-10$~eV in Ref. \cite{luschikov1969polarized}.
We obtained $P=0.250\pm0.050$ as the average value in the beamtime.
Fig.~\ref{fig:summary}(b) shows the neutron polarization $P_{\rm n}$ (white circle) and neutron transmission $T_{\rm n}$ (filled circle). 
\textcolor{black}{The neutron transmission $T_{\rm n}$ ($T_{\rm n0}$) is obtained by dividing the TOF spectrum of the polarized (unpolarized) naphthalene measurement by the TOF spectrum of the blank measurement. By substituting the transmission $T_{\rm n0}$ for Eq.~\ref{eq:trans_pol}, we obtained the cross section of naphthalene (see Fig. \ref{fig:summary}(a)). Here, the number density of the naphthalene is $5.36 \times 10^{21}$~$\rm /cm^3$.}

Fig.~\ref{fig:summary}(c) shows the FOM.
These figures of the neutron energy region from 1~meV to 10~eV are shown in Supplementary Fig.~S4.
Average values at $0.1-10$~eV were $P_{\rm n}=0.076\pm0.015$, $T_{\rm n}=0.555\pm0.001$, and ${\rm FOM}=0.0032\pm0.0009$. 
These results can be expected to be almost the same values up to keV because of the flat neutron-proton scattering cross-section.

The quality of the polarized proton spin filter was judged in comparison with the polarized $^{3}$ He one. 
The FOM of the polarized $^{3}$ He case is defined using the same discussion in Section \ref{sec:principle_spinfilter}.
The difference from the proton case is that the $^{3}$ He cross-section is dominated by the capture cross-section, such the spin-dependent cross-section is given by the same value but the opposite sign of the capture cross-section.
Assuming the performance of the polarized $^{3}$ He spin filter as $^{3}$ He polarization of 0.785 and a $^{3}$ He density of 41.5~bar$\cdot$cm~\cite{salhi2019situ}, we determined that the FOM of our spin filter exceeds it at 120~eV.
Further upgrades are necessary to achieve a higher FOM.

Increasing the thickness of the naphthalene crystal and the proton polarization is beneficial for the improvement of the FOM.
\textcolor{black}{Higher proton polarization} will be realized by cooling the chamber to a temperature lower than 90~K.
According to the previous study by PSI, to apply DNP to naphthalene crystals at a temperature of 25~K increases the proton spin-lattice relaxation time to 800~hrs and the proton polarization to 0.80 \cite{quan2019transportable}. 
Increasing the laser power and enlarging the microwave resonator is necessary for polarizing a thicker spin filter.
The laser intensity must be increased to irradiate the laser throughout the large filter.
The increase in laser intensity has the problem of exhausting the heat generated inside the crystal by absorbing the laser, but it can be cleared by the cooling.
Moreover, the enlargement of the microwave resonator means that the wavelength of the microwave becomes longer, which is equivalent to reducing the magnitude of the external magnetic field.
Since the cooling will result in a sufficiently long relaxation time, the reduction of the relaxation time by lowering the magnetic field is much smaller.
Therefore, all the necessary items for improving the FOM can be solved by lowering the temperature.

We will evaluate the performance of the improved spin filter.
It is also intended to confirm that the performances yield the same values as those at $0.1-10$~eV, even at higher neutron energies.
We plan to halve the pulse width of the RANS proton beam for evaluation at higher \textcolor{black}{energy}. 
If the pulse width is reduced by half, it can be used up to 40~eV, but the statistics will also be halved.
The lack of statistics due to the half pulse width is resolved by increasing the thickness of the ZnS (Li) scintillator of the RPMT detector.
The detection efficiency of the scintillator of the RPMT detector, as stated in Section \ref{sec:rans_experiment}, is about 3\% at 10~eV \cite{satoh2018development}.
Using a 1~mm thick scintillator increases the detection efficiency to about 8\%.
Even if the scintillator thickness is increased to 1~mm, the position resolution is sufficiently acceptable.


\section{Conclusions} \label{sec:conclusion}
We developed the novel triplet-DNP spin filter for the polarization of the epithermal neutrons.
Polarized epithermal neutrons are useful for studying compound states, which can answer open questions in physics.
The triplet-DNP spin filter has the advantage of a milder environment than that of the polarized proton spin filter based on conventional DNP.
Our spin filter, which is composed of a cylindrical naphthalene single crystal doped with 0.003 mol\% of pentacene-$d_{14}$, is the size of $\phi15\times4$~${\rm mm}^3$ and was operated at 0.35~T and 90~K. 
Performances of the triplet-DNP spin filter were evaluated using RANS. 
We succeeded in the neutron polarization of the epithermal region. 
The proton polarization was $0.250\pm0.050$ while the neutron polarization was $0.076\pm0.015$ at $0.1-10$~eV, respectively. 
The FOM of our spin filter exceeds that of the $^3$ He at $10^2$~eV. 
For our future study, we will develop an increasing size of the spin filter which can be operated below 90~K to improve the FOM.


\section*{Acknowledgements}
We especially thank the Advanced Manufacturing Support Team in RIKEN, who manufactured the microwave resonators.
We are grateful to Shunsuke Endo, Atsushi Kimura, Takashi Ino, and NOPTREX collaboration for their constructive discussions.
This work was supported by the RIKEN Incentive Research Projects.

\bibliographystyle{ptephy}
\bibliography{sample}

\end{document}